\documentclass[
    aps,
    prl,
    twocolumn,
    superscriptaddress,
    nofootinbib
]{revtex4-2}

\usepackage{amsmath,amssymb,amsfonts}
\usepackage{bm}
\usepackage{graphicx}
\usepackage{xcolor}
\usepackage[colorlinks=true,citecolor=blue,urlcolor=blue,linkcolor=blue]{hyperref}

\newcommand{\bk}{\bm{k}}

\begin{document}

\title{Complete Topological Classification with \(P\) and \(T\) Symmetries: Revealing a Topological Invariant Invisible to \(K\)-Theory}

\author{Chen Zhang}
\affiliation{Department of Physics and HK Institute of Quantum Science \& Technology, The University of Hong Kong, Pokfulam Road, Hong Kong, China}

\author{Y. X. Zhao}
\email{yuxinphy@hku.hk}
\affiliation{Department of Physics and HK Institute of Quantum Science \& Technology, The University of Hong Kong, Pokfulam Road, Hong Kong, China}

\begin{abstract}
The \(K\)-theoretic framework provides a complete topological classification of the tenfold symmetry classes and has been generalized to incorporate crystalline symmetries. Here, we show that this classification is incomplete even in the elementary case of spinless systems possessing both $P$ and $T$ symmetries. We obtain the complete classification through a first-principles analysis of the topological classes of \(P\)- and \(T\)-symmetric bands, namely, by classifying the corresponding symmetric clutching data. We identify a topological invariant that is invisible from the \(K\)-theoretic perspective when the occupied states at each inversion-invariant momentum have uniformly positive or uniformly negative parity. In special cases, such as when all inversion-invariant momenta have uniformly positive parity, this invariant can be interpreted as the second Stiefel--Whitney class or the Euler number defined over an inversion fundamental domain, namely, half of the Brillouin zone. Our work not only reveals a new topological invariant that cannot be determined from the parity spectra at inversion-invariant momenta under \(P\) and \(T\) symmetries, but also demonstrates the existence of crystalline topological phases that are absent from the \(K\)-theoretic classification.
\end{abstract}

\maketitle

{\color{blue}\textit{Introduction}} The tenfold periodic table for topological insulators and topological superconductors is a milestone in the theory of symmetry-protected topological phases~\cite{schnyder2008classification,kitaev2009periodic,ryu2010topological,chiu2016classification}. This classification also established the K-theoretical framework for the topological classification of symmetric band structures. The tenfold periodic table incorporates internal anti-unitary symmetries, including time-reversal symmetry and particle-hole symmetry, as well as their combination, known as chiral symmetry. The corresponding K theory is real K theory~\cite{atiyah1966k}. Crystalline symmetries were soon taken into consideration, beginning with twofold symmetries and subsequently extending to more complicated crystalline symmetries~\cite{fu2011topological,tanaka2012experimental,shiozaki2014topology,kruthoff2017topological,shiozaki2017topological,cornfeld2021tenfold}, which correspond to equivariant K theory~\cite{segal1968equivariant,atiyah1969equivariant}. With the inclusion of increasingly complex crystalline symmetries, anti-unitary symmetries, and chiral symmetries, a comprehensive classification formalism was eventually established, namely, twisted equivariant K theory~\cite{freed2013twisted}.

The K-theoretical framework is known to allow for an arbitrarily large number of energy bands. If the number of bands is restricted, exceptional topological invariants may exist that become unstable upon the addition of more bands, as exemplified by Hopf insulators and fragile crystalline topological insulators~\cite{moore2008topological,po2018fragile,liu2019shift}. It is therefore commonly believed in the community that, for a given set of symmetries, twisted equivariant real K theory provides a complete classification in the limit of a large number of bands.

In this work, we present a counterexample to this common belief. The symmetries under consideration are simple and elementary: spinless time-reversal symmetry $T$ and spatial inversion symmetry $P$.

The combined symmetry $PT$ has previously attracted considerable attention~\cite{zhao2016unified,zhao2017pt,ahn2018band,ahn2019failure,wu2019non,wang2020boundary,PhysRevLett.126.196402,chen2022second,xue2023stiefel}. This symmetry gives rise to a real band structure, whose complete classification is therefore provided by orthogonal K theory and characterized by the first and second Stiefel--Whitney classes~\cite{zhao2016unified,zhao2017pt,ahn2018band}. In particular, the novel bulk-boundary correspondence associated with the second Stiefel--Whitney class has been theoretically established and broadly explored in both quantum materials and artificial crystals~\cite{wang2020boundary,dai2021takagi,peng2022phonons,pan2022phononic,zhao2022quantum,chen2022second,pan2022two,zeng2023three,xue2023stiefel,zhang2023magnetic,gong2024hidden,wang2024mirror,yue2024stability,bouhon2024second,han2024crossed,wu2025breakdown,pal2025multi,li2025general,huang2025gapped,dai2025asymmetric,han2026real,kong2026efficient}.

In the presence of both $P$ and $T$, the valence states at each inversion-invariant momentum in the $2$D Brillouin zone are labeled by the parity eigenvalues $\pm 1$ due to inversion symmetry $P$. The parity spectra at all inversion-invariant momenta determine the Stiefel--Whitney classes~\cite{ahn2018band}. Furthermore, we show that the K-theoretical classification in the presence of both $P$ and $T$ is also completely characterized by the parity spectra; that is, it can be described by the theory of symmetry indicators~\cite{po2017symmetry,zhang2019catalogue,vergniory2019complete,tang2019comprehensive}.

We further provide a complete classification of band structures invariant under both $T$ and $P$ and identify a new $\mathbb{Z}_2$ invariant $\eta$ that is absent from the K-theoretical classification. This invariant exists if and only if the parity spectrum at each individual inversion-invariant momentum is either fully negative or fully positive. The simplest case occurs when all inversion-invariant momenta have fully positive parity. In this case, the Stiefel--Whitney classes are necessarily trivial, and the K-theoretical classification cannot distinguish the band structures. Nevertheless, $\eta$ reveals that the space of band structures has two distinct connected components. Moreover, because the number of valence bands can be arbitrarily large, $\eta$ is certainly not fragile in the conventional sense.

Interestingly, the condition for defining $\eta$ coincides with the condition under which the Wilson loop over half of the two-dimensional Brillouin zone can be used to formulate a topological invariant. This follows because $W(k_x)$, with $k_x\in[0,\pi]$, takes values in $O(N)$, while $W(0)$ and $W(\pi)$ are fixed at specified points $\pm \mathbb I_N$ in $O(N)$. We show that the spectral flow of the Wilson loop provides an explicit formulation of $\eta$.

To demonstrate that $\eta$ is not atomic but instead represents a physically topological phase, we investigate its bulk-boundary correspondence using the Dirac-model approach. Although the minimal Dirac model hosts helical edge states, we include all symmetry-allowed perturbations with random coefficients and find that the most stable phase is second order, with four corner states localized at the four corners of a rectangular sample. This behavior differs from that of a Stiefel--Whitney insulator, in which only one pair of corners hosts in-gap states.

{\color{blue}\textit{$P$-symmetric transition functions}} We consider a two-dimensional gapped spinless system with inversion symmetry \(P\) and time-reversal symmetry \(T\), satisfying
\begin{equation}
	[P,T]=0, \quad P^2=T^2=1.
\end{equation}
It is convenient to consider the combined symmetry \(PT\) together with \(P\). Since \(PT\) produces a real band structure, as discussed above, we only need to classify \(P\)-symmetric real band structures. In an appropriate basis, the \(PT\) operator is represented by \(\mathcal{K}\), the complex-conjugation operator. Accordingly, the Bloch Hamiltonian \(H(\bm{k})\) can be taken to be real, and the \(P\) symmetry is represented by a unitary matrix \(\mathcal{U}_P\):
\begin{equation}
	\mathcal U_P H(\bk)\mathcal U_P^\dagger
	=
	H(-\bk), \quad H(\bm{k})^*=H(\bm{k}).
	\label{eq:symmetries}
\end{equation}
In fact, \(\mathcal{U}_P\) is an orthogonal matrix because \([P,PT]=0\).

The idea behind the complete classification is to trivialize the real valence-band wave functions on two inversion-related contractible regions and then derive the clutching data along their boundaries under \(P\) symmetry. This yields a set of \(O(N)\)-valued transition functions subject to \(P\)-symmetric constraints. The complete classification is given by the path-connected components of this clutching-data space.

\begin{figure}[t]
	\centering
	\includegraphics[width=\linewidth]{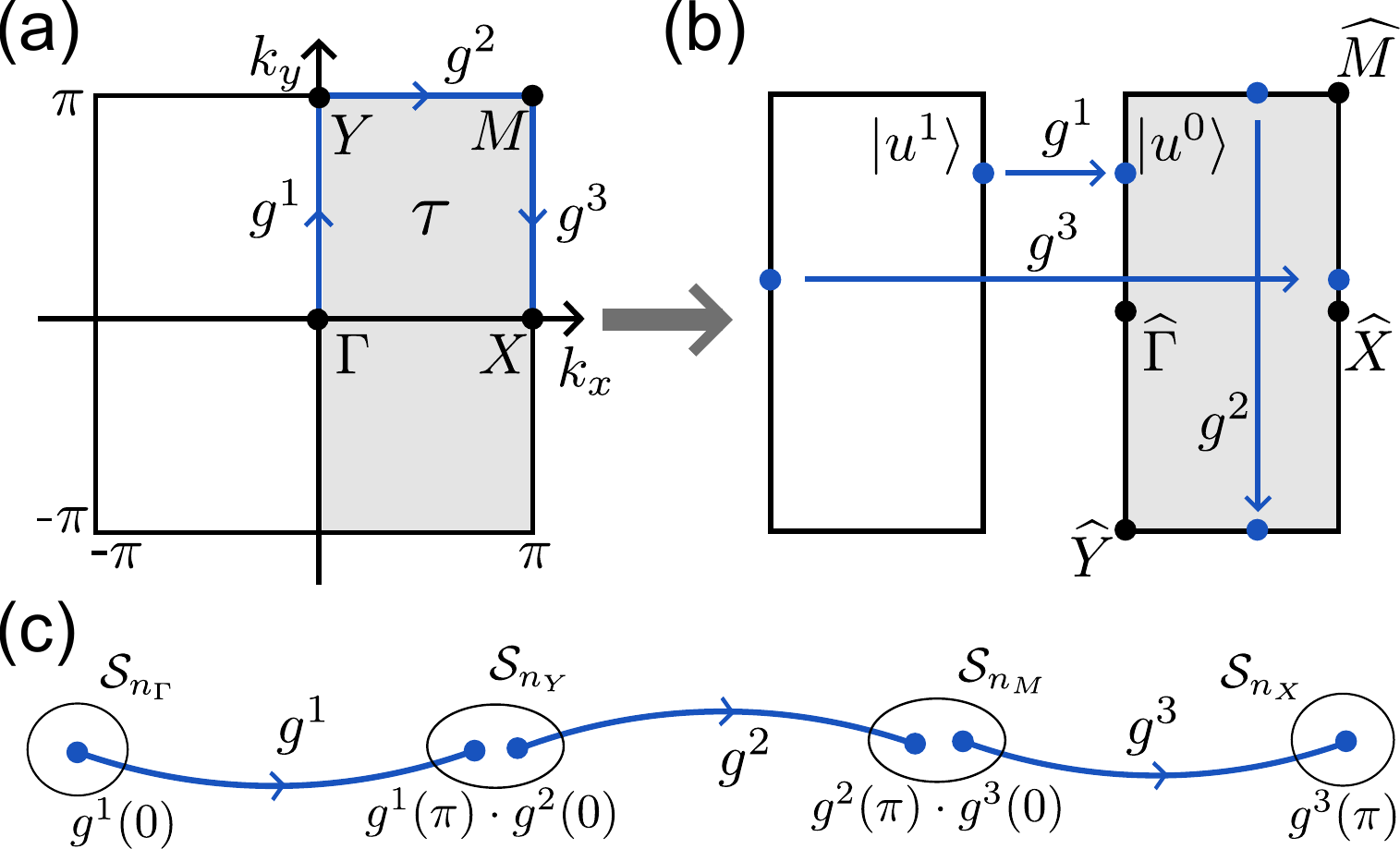}
	\caption{ \(P\)-symmetric transition functions.
		(a) The Brillouin torus is divided into an inversion fundamental domain \(\tau\) and its image \(P\tau\). The three blue paths \(g^1,g^2,g^3\) represent the independent transition functions.
		(b) After the torus is cut open, the two bases \(|u^0\rangle\) on \(\tau\) and \(|u^1\rangle\) on \(P\tau\) are glued along the boundaries by these transition functions.
		(c) The endpoints and junction products of the three successive paths are restricted to the subspace \(\mathcal{S}\) of symmetric orthogonal matrices.
	}
	\label{fig:clutching}
\end{figure}

To implement this construction, without loss of generality, we cut along \(k_x=0\) and divide the Brillouin torus into \(\tau=[0,\pi]\times[-\pi,\pi]\) and its inversion image \(P\tau\), as shown in Fig.~\ref{fig:clutching}(a). Since \(\tau\) is contractible, we can always choose a continuous orthonormal basis of occupied states \(\{|u_i^0(\bk)\rangle\}_{i=1}^N\) on this region. For each \(\bm{k}\in P\tau\), an orthonormal basis \(|u_i^1(\bk)\rangle\) is then determined by \(P\) symmetry:
\begin{equation}
	|u_i^1(\bk)\rangle
	=
	\mathcal U_P^\dagger |u_i^0(-\bk)\rangle,
	\label{eq:inversion-frame}
\end{equation}
where \(-\bm{k}\in\tau\).

On the boundary \(\partial\tau\) of \(\tau\), each state can be expanded in either of two bases, and the orthogonal transformation between them defines the transition functions. As illustrated in Fig.~\ref{fig:clutching}(a), there are three transition functions:
\begin{align}
	[g^1(k_y)]_{ij}
	&=
	\langle u_i^1(0,k_y)|u_j^0(0,k_y)\rangle,
	\nonumber\\
	[g^2(k_x)]_{ij}
	&=
	\langle u_i^0(k_x,\pi)|u_j^0(k_x,-\pi)\rangle,
	\nonumber\\
	[g^3(k_y)]_{ij}
	&=
	\langle u_i^1(-\pi,\pi-k_y)|u_j^0(\pi,\pi-k_y)\rangle ,
	\label{eq:transition-functions}
\end{align}
with \(0\leq k_x,k_y\leq\pi\). Here, \(g^1\) and \(g^3\) arise from \(P\) symmetry, whereas \(g^2\) arises from the reciprocal-lattice translation along the \(k_y\) direction. We adopt the convention that all three paths start at \(0\) and end at \(\pi\). For \(g^1\) (\(g^3\)) on the \(P\)-invariant line \(k_x=0\) (\(k_x=\pi\)), only the upper half with \(k_y\in[0,\pi]\) is independent, because the lower half is determined by inversion (see Fig.~\ref{fig:clutching}). The functions \(g^a\), with \(a=1,2,3\), can therefore be regarded as three paths in \(O(N)\) that start, connect, and end at the four \(P\)-invariant momenta \(\Lambda_i=\Gamma,Y,M,X\). Consequently, \(P\) symmetry imposes the constraints
\begin{multline}\label{eq:connection}
	g^1(0)^2=[g^1(\pi)g^2(0)]^2\\
	=[g^2(\pi)g^3(0)]^2=g^3(\pi)^2=\mathbb I_N.
\end{multline}
This follows because \(P\) is represented at the \(P\)-invariant momenta according to
\begin{equation}
	\begin{aligned}
		U_P(\Gamma)&=g^1(0),&
		U_P(Y)&=g^1(\pi)g^2(0),\\
		U_P(M)&=g^2(\pi)g^3(0),&
		U_P(X)&=g^3(\pi),
	\end{aligned}
	\label{eq:endpoint-constraints}
\end{equation}
where
\begin{equation}
[U_P(\Lambda_i)]_{mn}
=
\langle u_m^0(\widehat\Lambda_i)|
\mathcal U_P
|u_n^0(\widehat\Lambda_i)\rangle.
\end{equation}
The points \(\widehat{\Lambda}_i\) denote the chosen representatives in the region \(\tau\), as shown in Fig.~\ref{fig:clutching}(b).

Thus, the effect of \(P\) symmetry is to restrict each \(U_P(\Lambda_i)\) to the subspace \(\mathcal{S}\) of symmetric orthogonal matrices. A symmetric orthogonal matrix \(A\in O(N)\) satisfies \(A^2=\mathbb I_N\), and is therefore characterized by the degeneracy of its parity-\(-1\) eigenspace. The subspace \(\mathcal{S}\) has \(N+1\) path-connected components \(\mathcal{S}_n\), with \(n=0,1,\ldots,N\), distinguished by the degeneracy \(n\) of the eigenvalue \(-1\). Each component can be identified with the real Grassmannian
\begin{equation}
	\mathcal S_n\cong  O(N)/(O(n)\times O(N-n)).
\end{equation}
For \(N\geq3\), \(\pi_1(\mathcal S_n)\cong\mathbb Z_2\) when \(0<n<N\), whereas \(\mathcal S_0\) and \(\mathcal S_N\) are single points.

{\color{blue}\textit{The complete classification}} The classification problem has thus been reduced to determining the path-connected components of the space \(\mathcal{C}\) of three paths \(g^a\) in \(O(N)\) subject to the constraints in Eq.~\eqref{eq:connection}.

Let \(n_i\) denote the degeneracy of the \(P=-1\) eigenspace at \(\Lambda_i\). Not every \(\bm{n}\) is admissible in a \(P\)-invariant real band structure. The vector \(\bm{n}\) is admissible if and only if
\begin{equation}
	\sum_i n_i\equiv0\pmod2.
	\label{eq:parity-constraint}
\end{equation}
Each path \(g^a\) lies in one of the two connected components of \(O(N)\), determined by the constant value \(\det g^a=\pm1\). Thus,
\begin{equation}
	\begin{aligned}
		(-1)^{n_\Gamma}&=\det g^1,&
		(-1)^{n_Y}&=\det g^1\det g^2,\\
		(-1)^{n_M}&=\det g^2\det g^3,&
		(-1)^{n_X}&=\det g^3,
	\end{aligned}
	\label{eq:determinant-constraints}
\end{equation}
and Eq.~\eqref{eq:parity-constraint} follows by multiplying these four equations.

For each admissible \(\bm{n}\), let \(\mathcal{C}_{\bm{n}}\) denote the subspace of \(\mathcal{C}\) in which \(U_P(\Lambda_i)\) is restricted to \(\mathcal{S}_{n_i}\) for each \(i\). Clearly, different \(\bm{n}\) correspond to distinct path-connected components of \(\mathcal{C}\). The remaining task is therefore to identify the path-connected components \(\pi_0(\mathcal{C}_{\bm{n}})\) for each admissible \(\bm{n}\). Applying the theorem in Appendix~B gives
\begin{equation}
	\pi_0(\mathcal C_{\bm n})
	\cong
	\frac{\pi_1(O(N))}
	{H_{n_\Gamma}+H_{n_Y}+H_{n_M}+H_{n_X}},
	\label{eq:path-space-quotient}
\end{equation}
as sets. Here, $
H_n=
\operatorname{im}\!\left[
\iota_*:\pi_1(\mathcal S_n)\to \pi_1(O(N))
\right]$,
where \(\iota\) denotes the inclusion of \(\mathcal{S}_n\) into \(O(N)\). To understand this result, it is useful to recall the following fact. For a path-connected space \(X\), let \(\mathcal{C}_{a,B}\) be the space of paths that start at \(a\in X\) and end in a subspace \(B\subset X\). Then
$
\pi_0(\mathcal{C}_{a,B})\cong \pi_1(X)/\iota_*\pi_1(B).
$

For \(0<n<N\), the inclusion \(\iota:\mathcal S_n\hookrightarrow O(N)\) induces a surjective map \(\iota_*\), and hence \(H_n\cong\pi_1(O(N))\). Therefore, a single \(n_i\) satisfying \(0<n_i<N\) is sufficient to make \(\pi_0(\mathcal{C}_{\bm n})\) trivial. An additional invariant exists only when \(n_i\in\{0,N\}\) for all \(i\), in which case all \(\mathcal S_{n_i}\) are single points and \(H_{n_i}=0\). In this case, there is an additional invariant \(\eta\), because
\begin{equation}
	\pi_0(\mathcal C_{\bm n})
	\cong \pi_1(O(N))
\end{equation}
as sets.

In conclusion, the complete classification is characterized by
\begin{equation}
	\bigl(n_\Gamma,n_Y,n_M,n_X;\eta\bigr),
	\label{eq:complete-classification}
\end{equation}
where \(\bm{n}\) satisfies Eq.~\eqref{eq:parity-constraint}. The invariant \(\eta\) takes values in \(\pi_1(O(N))\) when \(n_i\in\{0,N\}\) for all \(i\), and otherwise \(\eta=0\). Recall that \(\pi_1(O(N))\cong\mathbb{Z}_2\) for \(N\geq3\), and \(\pi_1(O(2))\cong\mathbb{Z}\).

\begin{figure}[t]
	\centering
	\includegraphics[width=\linewidth]{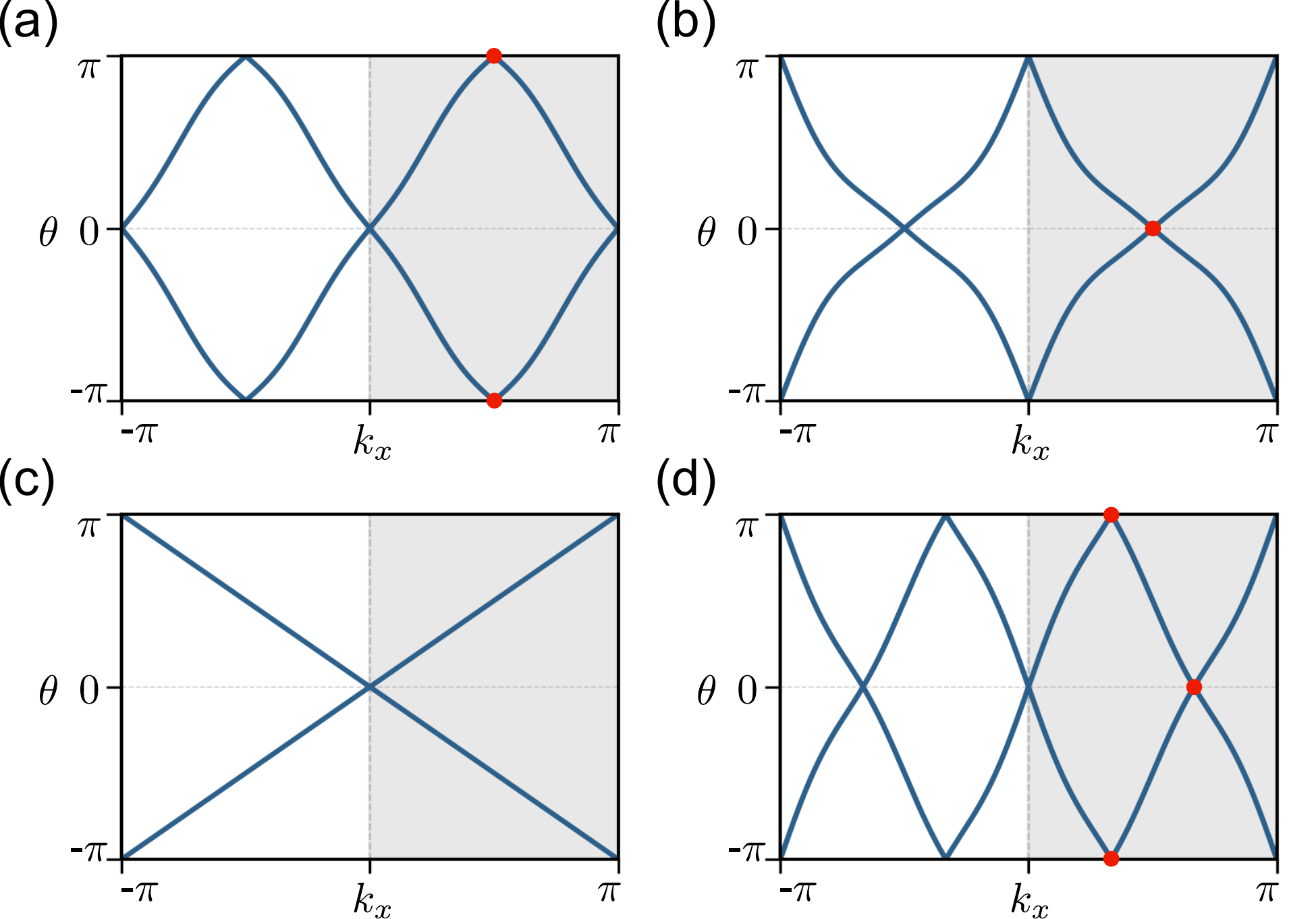}
	\caption{
	Schematic Wilson-loop spectral flows for the \(\eta\) invariant. The half Brillouin zone \(\tau\) is shaded.
	(a) A single winding with both endpoints pinned at \(\theta=0\).
	(b) A single winding with both endpoints pinned at \(\theta=\pi\).
	(c) The reference spectral flow with the two endpoints pinned at \(\theta=0\) and \(\theta=\pi\), respectively.
	(d) A single winding relative to the reference spectral flow in (c).
	}
	\label{fig:wilson-schematic}
\end{figure}

{\color{blue}\textit{The Wilson-loop method}} We now explain how to compute \(\eta\) using Wilson-loop spectral flow when an admissible \(\bm{n}\) satisfies \(n_i\in\{0,N\}\) for all \(i\).

For a smooth orthonormal basis on \(\tau\), the Wilson-loop operator \(W(k_x)\) for each \(k_y\) subsystem, parametrized by \(k_x\in[0,\pi]\), is given by
\begin{equation}
	W(k_x)=\mathcal P\exp\!\left[-i\!\int_{-\pi}^{\pi}dk_y\,
	A_y(k_x,k_y)\right]g^2(k_x),
	\label{eq:wilson-definition}
\end{equation}
where $[A_y(k_x,k_y)]_{ij}=\langle u_i^0(k_x,k_y)|i\partial_{k_y}|u_j^0(k_x,k_y)\rangle$.
The Wilson-loop operator \(W(k_x)\) can be regarded as a path in \(O(N)\). Such a path can be topologically nontrivial only when its endpoints are restricted, as in the mathematical fact recalled below Eq.~\eqref{eq:path-space-quotient}.

This restriction occurs precisely when \(n_i\in\{0,N\}\) for all \(i\), namely when
\begin{equation}
U_P(\Lambda_i)=(-1)^{s_i}\mathbb I_N,
\qquad s_i=0,1.
\end{equation}
It is then straightforward to derive
\begin{equation}
	\begin{aligned}
		W(0)&=U_P(\Gamma)U_P(Y)=(-1)^{s_\Gamma+s_Y}\mathbb I_N,\\
		W(\pi)&=U_P(M)U_P(X)=(-1)^{s_M+s_X}\mathbb I_N .
	\end{aligned}
	\label{eq:wilson-endpoints}
\end{equation}
With the starting and ending points pinned at \(\pm\mathbb I_N\), the path-connected components of the path space are in bijection with \(\pi_1(O(N))\). In Appendix~C, we show that the component of the path \(W(k_x)\) corresponds precisely to the invariant \(\eta\).

The component of \(W(k_x)\) can be determined from its spectral flow. Let \(\{e^{i\theta_n(k_x)}\}\) be the eigenvalues of \(W(k_x)\). Since \(W(k_x)\in O(N)\), its eigenvalues occur in conjugate pairs of \(U(1)\) phases. Generically, with the endpoints pinned, each pair winds around \(U(1)\) as \(k_x\) runs from \(0\) to \(\pi\), and the sum of the winding numbers over all pairs determines the component of \(W(k_x)\), reduced modulo \(2\) when \(N\geq3\).

If \(W(0)=W(\pi)=\mathbb{I}_N\) [or \(W(0)=W(\pi)=-\mathbb{I}_N\)], the winding number of a pair is given by the number of times the spectral flow crosses \(\theta=\pi\) [or \(\theta=0\)], as illustrated in Fig.~\ref{fig:wilson-schematic}(a) [or Fig.~\ref{fig:wilson-schematic}(b)]. If \(W(0)=\mathbb{I}_N\) and \(W(\pi)=-\mathbb{I}_N\), or vice versa, we can take the straight path as a reference path. A complete winding must then cross both \(\theta=0\) and \(\theta=\pi\) once in the interior of the path, as illustrated in Fig.~\ref{fig:wilson-schematic}(c) and (d).

It is noteworthy that, over the half Brillouin zone \(\tau\), the invariant \(\eta\), expressed in terms of Wilson-loop spectral flow, resembles the second Stiefel--Whitney number for \(N\geq3\) or the Euler number for \(N=2\) defined over the full Brillouin zone. 

{\color{blue}\textit{Model realization and boundary response}} We now construct a Dirac model realizing a nontrivial \(\eta\) invariant in the case where \(U_P(\Lambda_i)=\mathbb{I}_N\) for all \(i\).

Using two sets of Pauli matrices, \(\sigma\) and \(\tau\), we choose the Dirac gamma matrices as
$\gamma^1=\sigma_3\otimes\tau_1$,
$\gamma^2=\sigma_3\otimes\tau_3$,
$\gamma^3=\sigma_1\otimes\tau_0$,
$\gamma^4=\sigma_3\otimes\tau_2$ and $\gamma^5=\sigma_2\otimes\tau_0$, satisfying $\{\gamma^\mu,\gamma^\nu\}=2\delta^{\mu\nu}\mathbb{I}_4$. Note that $\gamma^{1,2,3}$ are real, whereas \(\gamma^{4,5}\) are imaginary. The minimal model is constructed as
\begin{equation}
	H_0(\bk)=a(\bk)\gamma^1+b(\bk)\gamma^2+c(\bk)\gamma^3,
	\label{eq:model}
\end{equation}
with $a(\bk)=\cos2k_y-\cos2k_x$, $b(\bk)=2\sin k_x\sin k_y$, $c(\bk)=t-\cos k_x-\cos k_y$.
The Hamiltonian is real and even under \(\bk\mapsto-\bk\), so it has \(T=\mathcal K\hat I\) and \(P=\mathbb I_4\hat I\). Its spectrum is
$E_\pm(\bk)=\pm\sqrt{a^2+b^2+c^2}$,
with twofold degeneracy. For the two valence bands, \(U_P(\Lambda_i)=\mathbb I_2\) by construction. Although \(N=2\) in this model, the parity of the winding number corresponds to the \(\eta\) invariant after trivial valence bands with \(P=1\) at each \(\Lambda_i\) are added. Moreover, all topological features are stable under the symmetry-preserving constant perturbations
\begin{equation}
	\Delta H=i\gamma^1(m_1\gamma^4+m_2\gamma^5)
	+i\gamma^3(m_3\gamma^4+m_4\gamma^5).
	\label{eq:perturbation}
\end{equation}

\begin{figure}[t]
	\centering
	\includegraphics[width=\linewidth]{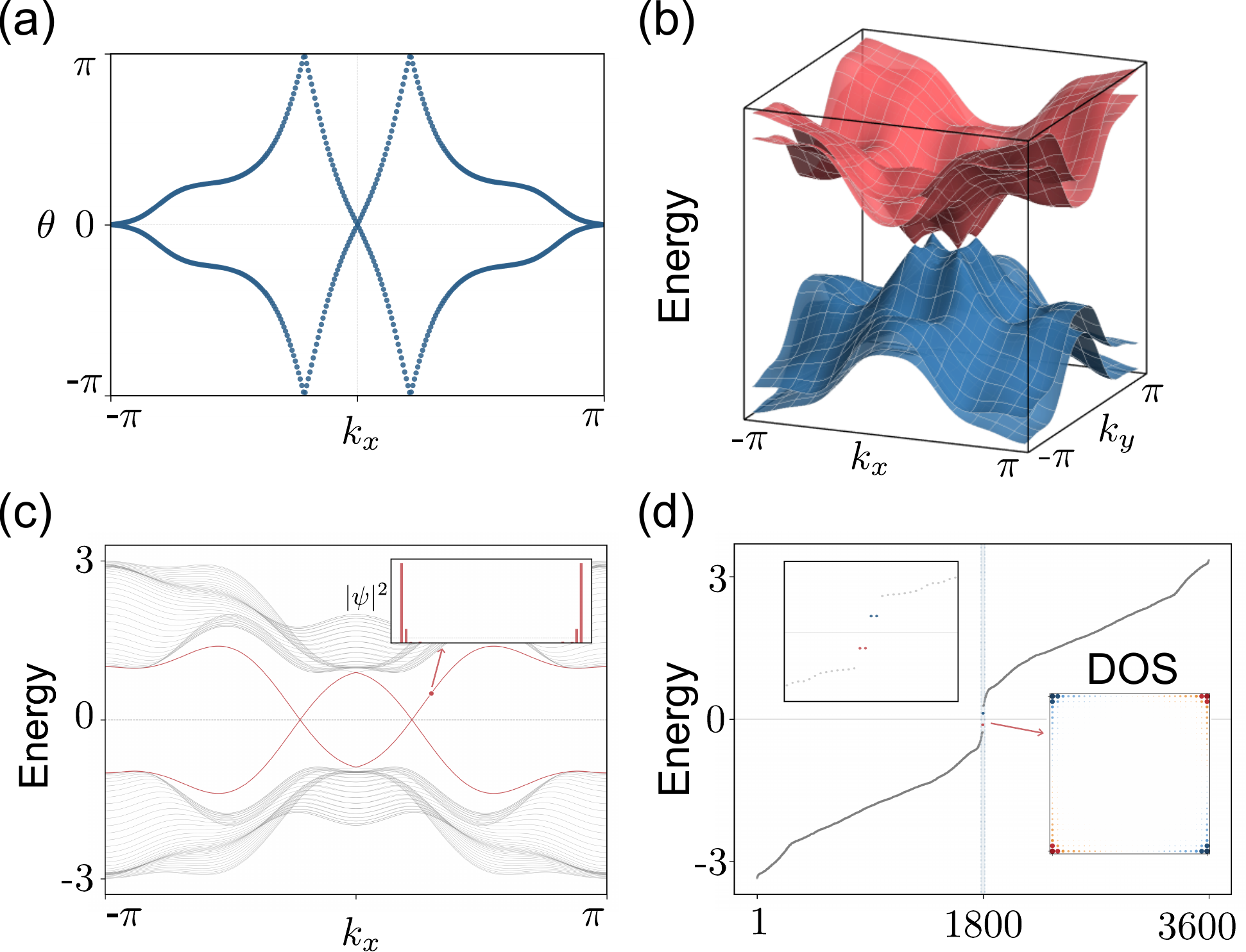}
	\caption{The Dirac model. (a) The Wilson-loop spectral flow with \(t=1\).  The spectral flow over the half-zone exhibits a single crossing at \(\theta=\pi\), yielding \(\eta=1\).
	(b) Bulk spectrum at the phase transition for \(t=1.75\) and \((m_1,m_2,m_3,m_4)=(0.25,0.25,0.2,0.2)\).
	(c) The spectrum at \(t=1\) on a slab geometry with open boundary conditions along the \(y\) direction.
	(d) Spectrum of a finite square at \(t=1\), with symmetry-preserving boundary perturbations \((m_1,m_2,m_3,m_4)=(0.25,0.25,0.2,-0.2)\).
	The four in-gap states are localized at corners, as illustrated by the DOS inset.
	}
	\label{fig:model}
\end{figure}

Figure~\ref{fig:model}(a) shows the Wilson-loop spectral flow in the nontrivial phase with \(t=1\). The Wilson phases cross \(\theta=\pi\) once for \(0<k_x<\pi\), giving \(\eta=1\). Figure~\ref{fig:model}(b) shows the topological phase transition, in which a symmetry-preserving perturbation splits the quadratic gap-closing point at \(\Gamma\) into four Dirac points.

The nontrivial \(\eta\) invariant corresponds to boundary states and is therefore not atomic. For a slab geometry with open boundary conditions in one dimension, the spectrum contains two in-gap helical modes on each edge, as shown in Fig.~\ref{fig:model}(c). These helical modes constitute a critical point of the boundary-state spectrum, since the symmetry-preserving perturbation in Eq.~\eqref{eq:perturbation} can gap the edge. On a rectangular geometry, however, four in-gap states remain, one at each corner, and are stable under random perturbations of the form in Eq.~\eqref{eq:perturbation}. Recall that a Stiefel--Whitney number can produce a critical helical mode and a pair of \(PT\)-related corner states~\cite{wang2020boundary}. The resemblance between \(\eta\) and a Stiefel--Whitney or Euler number defined over the half Brillouin zone \(\tau\) may naturally explain the observed bulk-boundary correspondence: inversion symmetry protects the coexistence of two Stiefel--Whitney or Euler bulk-boundary correspondences.

{\color{blue}\textit{Summary and discussions}} 
In summary, we have established a complete topological classification of spinless systems invariant under both inversion and time-reversal symmetries. In doing so, we have identified a new topological invariant, \(\eta\), that is invisible from the perspective of \(K\)-theoretic classification.

This limitation can be understood from the structure of equivariant \(K\)-groups. In the presence of a fixed point, the topological information is encoded in the reduced \(K\)-group, which is unchanged by the addition of flat bands induced from representations of the symmetry group. In our case, one may freely add flat bands carrying the same arbitrary representation of inversion symmetry at all inversion-invariant momenta without changing the reduced \(K\)-theory class. Such an addition, however, can violate the condition required for the invariant \(\eta\), namely that the valence states at each inversion-invariant momentum must have either uniformly positive or uniformly negative parity.

Our results demonstrate that \(K\)-theoretic classification does not, in general, provide a complete classification of crystalline topological phases, even though the tenfold-way periodic table is complete for the corresponding internal symmetries. A first-principles analysis is therefore generally necessary, as illustrated here by the emergence of a new topological invariant in a system with these elementary symmetries.

\begin{acknowledgments}
This work was supported by the Research Grants
Council of Hong Kong through the General Research
Fund (Grant Nos. 17301224 and 17302525).
\end{acknowledgments}

\vskip 12pt
\appendix 
{\color{blue}\textit{Appendix A on the equivariant \(KO\)-group.}} The corresponding K-theoretical classification of two-dimensional insulators invariant under both $P$ and $T$ is given by the equivariant orthogonal K-group \(KO_{C_2}(T^2)\), where \(C_2\) is the point group generated by $P$ and acting on real vector bundles over the two-torus $T^2$.

We view the inversion-symmetric Brillouin torus as a product of two inversion-symmetric circles,
\begin{equation}
	T^2= S^1\times S^1 .
\end{equation}
Here, \(S^1\) denotes a Brillouin circle with the natural inversion action. Both $T^2$ and \(S^1\) have a natural base point, namely, the $\Gamma$ point, or Brillouin-zone center. For any two based spaces (CW complexes) $X$ and $Y$, there is a homotopy equivalence
\begin{equation}
\Sigma (X\times Y)\simeq \Sigma X \vee \Sigma Y \vee \Sigma(X\wedge Y),
\end{equation}
where $\Sigma$ denotes the suspension operator and $X\wedge Y$ denotes the smash product of $X$ and $Y$. Specializing to the present case, we obtain
\begin{equation}
	\Sigma T^2\simeq \Sigma S^1 \vee \Sigma S^1 \vee \Sigma S^2.
\end{equation}
Here, \(S^1\wedge S^1\cong S^2\), which can be regarded as the standard sphere on which $P$ acts as a $\pi$ rotation about the $z$ axis. Importantly, this homotopy equivalence preserves the inversion symmetry.

Let us recall the wedge axiom for reduced equivariant orthogonal K-theory~\cite{segal1968equivariant,may1996equivariant},
\begin{equation}
\widetilde{KO}^n_{G} \left(\bigvee_\lambda X_\lambda\right)\cong \prod_\lambda \widetilde{KO}^n_G(X_\lambda),
\end{equation}
for a finite wedge, together with the suspension isomorphism
\begin{equation}
    \widetilde{KO}_G^{n+1}(\Sigma X)\cong \widetilde{KO}_G^n(X).
\end{equation}
In the present case, these two axioms imply
\begin{equation}
	\widetilde{KO}_{C_2} (T^2)\cong  \widetilde{KO}_{C_2}(S^1)\oplus \widetilde{KO}_{C_2}(S^1) \oplus \widetilde{KO}_{C_2}(S^2),
\end{equation}
where $\widetilde{KO}_G(X)=\widetilde{KO}_G^{0}(X)$. It is a standard result that
\begin{equation}
\widetilde{KO}_{C_2}(S^1)\cong \widetilde{KO}_{C_2}(S^2)\cong \mathbb{Z}.
\end{equation}
These groups can be derived from the long exact sequences in equivariant orthogonal K-theory obtained by symmetrically collapsing the boundary of a disk $D^d$ to a single fixed point to form $S^d$, for $d=1$ and $2$.

Thus, we obtain $\widetilde{KO}_{C_2} (T^2)\cong \mathbb{Z}^3$ and therefore
\begin{equation}
	KO_{C_2}(T^2)\cong \mathbb{Z}^5.
\end{equation}
The additional two $\mathbb{Z}$ components are generated by the two irreducible representations of $C_2$, corresponding to $P=\pm 1$, respectively. Note that, for a space $X$ with a $G$-fixed point,
\begin{equation}
KO_G(X)=RO_G\oplus \widetilde{KO}_G(X),
\end{equation}
where $RO_G$ is the free Abelian group generated by the irreducible orthogonal representations of $G$.

The complete set of topological invariants for $KO_{C_2}(T^2)\cong \mathbb{Z}^5$ can be written as
\begin{equation}
	(N, n_\Gamma ; v_X, v_Y, m),
\end{equation}
where $v_X=n_X-n_\Gamma$, $v_Y=n_Y-n_\Gamma$, and
\begin{equation}
m=(n_\Gamma+n_M-n_X-n_Y)/2.
\end{equation}
The quantity $m$ is well defined because the admissible $n_i$ satisfy $\sum_i n_i\equiv0\pmod2$. These integers are all independent. The quantities $N$ and $n_\Gamma$ account for $RO_{C_2}$, since $N-n_\Gamma$ and $n_\Gamma$ correspond to $P=\pm 1$, respectively, at the $\Gamma$ point. The remaining three components characterize $\widetilde{KO}_{C_2}(T^2)$, because they are invariant under adding the same representation at all four inversion-invariant momenta.

{\color{blue}\textit{Appendix B on the complete classification.}}
In this section, we first prove the theorem in Eq.~\eqref{eq:path-space-quotient} and then discuss the image of $\pi_1(S_{n})$ in $\pi_1(O(N))$.

For fixed admissible \(\bm n=(n_\Gamma,n_Y,n_M,n_X)\), a triple
\((g^1,g^2,g^3)\in\mathcal C_{\bm n}\) can be folded into a single path,
\begin{equation}
	\mathcal G
	=
	g^1\ast\bigl[
	U_P(Y)\bigl(g^2\ast[U_P(M)(g^3)^{-1}]\bigr)^{-1}
	\bigr].
	\label{eq:folded-path}
\end{equation}
Here, \(\ast\) denotes path concatenation, and the inverse denotes pointwise group inversion. The junction conditions ensure that successive path segments match, so that the endpoint of each segment coincides with the initial point of the next one. Hence, \(\mathcal G\) is continuous in $O(N)$. Its initial and final points are
\begin{equation}
	\mathcal G(0)=U_P(\Gamma),
	\quad
	\mathcal G(1)=U_P(Y)U_P(X)U_P(M)^{-1},
\end{equation}
where the total time has been rescaled to $1$.

The final endpoint is parametrized by the product map
\begin{equation}
	\begin{aligned}
		F \colon
		\mathcal S_{n_Y}\times
		\mathcal S_{n_M}\times
		\mathcal S_{n_X}
		&\longrightarrow O(N),\\
		(A_Y,A_M,A_X)
		&\longmapsto A_YA_XA_M^{-1}.
	\end{aligned}
	\label{eq:endpoint-product-map}
\end{equation}
Thus, we have defined a path space $\mathcal{C}_{\mathcal{S}_{n_\Gamma}, F}$ consisting of paths whose initial points lie in $\mathcal S_{n_\Gamma}$ and whose final points are parametrized by $F_{\bm n}$. This path space is homeomorphic to $\mathcal{C}_{\bm{n}}$. Moreover, $\mathcal{C}_{\mathcal{S}_{n_\Gamma}, F}$ is nonempty if and only if the initial and final points lie in the same connected component of \(O(N)\), that is,
\begin{equation}
\det \mathcal G(0)=\det\mathcal G(1),
\end{equation}
which is equivalent to the condition \eqref{eq:parity-constraint} for admissible $\bm{n}$.

More generally, consider paths in a path-connected space $X$, with initial points restricted to a subspace $A$ and final points parametrized by a function $f:Y\rightarrow X$, where $Y$ is path-connected. Denote the corresponding path space by $\mathcal{C}_{A,f}$. Then,
\begin{equation}
	\pi_0(\mathcal{C}_{A,f})\cong H_A\backslash \pi_1(X)/\mathrm{im}f_*.
\end{equation}
Here, $H_A$ is the image of $\pi_1(A)$ in $\pi_1(X)$, and $f_*$ is the map induced from $\pi_1(Y)$ to $\pi_1(X)$. This result follows from the long exact sequences associated with $A\subset X$ and $f:Y\rightarrow X$ in relative homotopy theory~\cite{whitehead1948operators,hatcher2002algebraic}. If $\pi_1(X)$ is Abelian, then
\begin{equation}
	\pi_0(\mathcal{C}_{A,f})\cong \frac{\pi_1(X)}{H_A+\mathrm{im}f_*}.
\end{equation}

In the present problem, the initial points are restricted to \(\mathcal S_{n_\Gamma}\), while the final points are parametrized by \(F\). Therefore,
\begin{equation}
	\pi_0(\mathcal{C}_{\mathcal{S}_{n_\Gamma}, F})\cong \frac{\pi_1(O(N))}{H_{n_\Gamma}+\mathrm{im} F_*} .
\end{equation}
The fundamental groups of the two components of $O(N)$ are isomorphic. Since $O(N)$ is a Lie group, we have
\begin{equation}
	\operatorname{im}F_*
	=
	H_{n_Y}+H_{n_M}+H_{n_X}.
\end{equation}
Consequently,
\begin{equation}
	\pi_0(\mathcal{C}_{\mathcal{S}_{n_\Gamma}, F})
	\cong
	\frac{\pi_1(O(N))}
	{H_{n_\Gamma}+H_{n_Y}+H_{n_M}+H_{n_X}}.
	\label{eq:app-relative-quotient}
\end{equation}
Since $\mathcal{C}_{\bm{n}}\cong \mathcal{C}_{\mathcal{S}_{n_\Gamma}, F}$, this proves Eq.~\eqref{eq:path-space-quotient}.

It remains to determine \(H_n\). The spaces \(\mathcal S_0\) and $\mathcal{S}_N$ consist of the single points \(\mathbb I_N\) and \(-\mathbb I_N\), respectively, and hence \(H_0=H_N=0\). For \(0<n<N\), choose one positive-parity and one negative-parity direction, and let \(R(\theta)\) rotate the resulting two-dimensional plane. It is useful to observe that the path
\begin{equation}
	A_n(\theta)=R(\theta)D_nR(\theta)^{-1},
	\qquad 0\leq\theta\leq\pi,
	\label{eq:app-endpoint-loop}
\end{equation}
is a closed loop in \(\mathcal S_n\), where \(D_n=\operatorname{diag}(\mathbb I_{N-n},-\mathbb I_n)\). To see this, note that $D_nR(\theta)^{-1}=R(\theta)D_n$, and therefore
\begin{equation}
A_n(\theta)=R(2\theta)D_n.
\end{equation}
Since $R(2\theta)$ executes a full \(2\pi\) rotation, \(A_n(\theta)\) generates \(\pi_1(O(N))\).

It follows that
\begin{equation}
	H_n=
	\begin{cases}
		0,& n=0\ \text{or}\ n=N,\\
		\pi_1(O(N)),& 0<n<N.
	\end{cases}
	\label{eq:app-Hn-result}
\end{equation}
Consequently, if at least one \(n_i\) is intermediate, \(0<n_i<N\), the denominator in Eq.~\eqref{eq:app-relative-quotient} equals \(\pi_1(O(N))\), and \(\mathcal C_{\bm n}\) is connected. Otherwise, \(n_i\in\{0,N\}\) for all four inversion-invariant momenta, all \(H_{n_i}\) vanish, and hence
\begin{equation}
	\pi_0(\mathcal C_{\bm n})
	\cong \pi_1(O(N)).
\end{equation}

{\color{blue}\textit{Appendix C on the Wilson-loop method.}}
We consider a parallel-transport gauge along $\tau$, namely, we choose the basis
\begin{equation}
	|\widetilde u_i^0(\bm{k})\rangle=\sum_j
	|u_j^0(\bm{k})\rangle \left[
	\mathcal P e^{-i\int_{-\pi}^{k_y}dq_y\,A_y(k_x,q_y)}
	\right]_{ji}^{\dagger} ,
	\label{eq:app-parallel-frame}
\end{equation}
under which
\begin{equation}
	W(k_x)=\widetilde g^2(k_x).
	\label{eq:app-W-g2}
\end{equation}
Here, \(\widetilde{g}^a\) denotes the transition functions in the parallel-transport gauge.

We can also apply $P$ symmetry to obtain \( |\widetilde{u}_i^1\rangle\), which satisfies the parallel-transport gauge along \(P\tau\). These states satisfy
\begin{equation}
	\langle\widetilde u_i^0|\partial_{k_y}\widetilde u_j^0\rangle=\langle\widetilde u_i^1|\partial_{k_y}\widetilde u_j^1\rangle=0.
\end{equation}
Consequently, both \(\widetilde{g}^1\) and \(\widetilde{g}^3\) are constant. For \(\widetilde{g}^1\), for example,
\begin{equation*}
	\begin{split}
		&\partial_{k_y}\widetilde g^1_{ij}(k_y)\\
		=&
		\langle \partial_{k_y}\widetilde u_i^1|
		\tilde u_j^0\rangle
		+
		\langle \widetilde u_i^1|
		\partial_{k_y}\widetilde u_j^0\rangle \\
		=&
		\sum_m
		\langle \partial_{k_y}\widetilde u_i^1|\widetilde u_m^1\rangle
		\langle \widetilde u_m^1|\widetilde u_j^0\rangle
		+
		\sum_m
		\langle \widetilde u_i^1|\tilde u_m^0\rangle
		\langle \widetilde u_m^0|\partial_{k_y}\widetilde u_j^0\rangle 
		= 0 ,
	\end{split}
\end{equation*}
and similarly for \(\widetilde g^3\).

Thus, the homotopy class of the transition functions, which is gauge independent, is determined solely by \(\widetilde{g}^2\), and hence solely by the Wilson-loop operator $W(k_x)$ through Eq.~\eqref{eq:app-W-g2}. In other words, there is a bijection between the path-connected components of the \(g^a\) and the homotopy classes of \(W(k_x)\) viewed as a path in \(O(N)\). As discussed in the main text, \(W(k_x)\) has nontrivial homotopy classes precisely when all inversion-invariant momenta have fully positive or fully negative parity.

\bibliographystyle{apsrev4-2}
\bibliography{references}

\end{document}